\documentstyle[epsfig]{mn}
\begin{document}
\def\ltsima{$\; \buildrel < \over \sim \;$}
\def\simlt{\lower.5ex\hbox{\ltsima}}
\def\gtsima{$\; \buildrel > \over \sim \;$}
\def\simgt{\lower.5ex\hbox{\gtsima}}
\def\approxgt{\mathrel{\hbox{\rlap{\lower.55ex \hbox {$\sim$}}
        \kern-.3em \raise.4ex \hbox{$>$}}}}
\def\approxlt{\mathrel{\hbox{\rlap{\lower.55ex \hbox {$\sim$}}
        \kern-.3em \raise.4ex \hbox{$<$}}}}

\def \src {IGR~J16479-4514}
 
\def \swift {\emph{Swift}}
\def \int {$INTEGRAL$}
\def \xmm {$XMM$-$Newton$}
\def \rosat {$ROSAT$}
\def \asca {$ASCA$}
\def \sw {$Swift$}
\def \xte {$RXTE$}
\def \chandra {$Chandra$}
\def \suz {$Suzaku$}

\title[]{Chasing candidate Supergiant Fast X-ray Transients in the 1,000 orbits \int/IBIS catalog}  

\author[Sguera et al.]
{V. Sguera$^{1}$, L. Sidoli$^2$,  A. J. Bird$^3$, A. Paizis$^2$,  A. Bazzano$^4$ \\ 
$^1$  INAF$-$OAS, Osservatorio di Astrofisica e Scienza dello Spazio, Area della Ricerca del CNR, via Gobetti 101, I-1-40129 Bologna, Italy \\
$^2$ INAF$-$IASF, Istituto di Astrofisica Spaziale e Fisica Cosmica, Via A. Corti   12, I-20133 Milano, Italy \\
$^3$ School of Physics and Astronomy, Faculty of Physical Sciences and Engineering, University of Southampton, SO17 1BJ, UK \\
$^4$ INAF$-$IAPS, Istituto di Astrofisica e Planetologia Spaziali, Via Fosso del Cavaliere 100, I-00133 Roma, Italy \\
}

\date{Accepted 2019 November 21. In original form 2019 October 15}

\maketitle

\begin{abstract}
We report results from an investigation at hard X-rays (above 18 keV) and soft X-rays (below 10 keV) of a sample of  X-ray transients located on the Galactic plane and detected with the bursticity method, as reported in the latest 1,000 orbits \int/IBIS catalog. Our  main aim has been to  individuate  those with  
X-rays characteristics strongly resembling  Supergiant  Fast X-ray Transients (SFXTs).  As a result, we found four unidentified fast X-ray transients which now can be considered good SFXT candidates. In particular,  three transients (IGR~J16374$-$5043, IGR~J17375$-$3022 and IGR~J12341$-$6143) were very poorly studied in the literature before the current work, and our findings largely improved the knowledge of their X-ray characteristics. 
The other transient (XTE~J1829$-$098) was previously  studied in detail only below 10 keV, conversely the current work provides the first detailed study in outburst above 18 keV. In addition we used archival infrared observations of the transients to pinpoint, among the field objects, their best candidate counterpart. We found that their photometric properties are compatible with an early type spectral classification,  further  supporting our proposed nature of SFXTs. Infrared spectroscopy is advised to confirm or disprove our interpretation. The reported findings allowed a  significant increase of  the sample of candidate SFXTs known to date, effectively doubling their number. 
\end{abstract}


\vspace{1.0cm}

\section{Introduction}

The X-ray sky is extremely variable, markedly characterized by many transient sources which suddenly turn on and 
then disappear  on short (hours) or long (weeks, months) timescales.  Especially above 20 keV,   such variability has not 
been fully exploited yet since  to date hard X-ray surveys have mostly explored the persistent sky.
Indeed several \int/IBIS and  \swift/BAT hard X-ray catalogs have been published in the literature, 
mainly devoted to search for persistent sources  detected by accumulating as much exposure as possible 
(e.g. Bird et al. 2006, 2007, 2010, Kyuseok et al. 2018, Krivonos et al. 2017, Baumgartner et al. 2013). 
As a drawback, this approach could eventually lead to difficulties  in detecting transient objects. For example,  a transient  found in early catalogues may  have, after long periods of quiescence, an undetectable low mean flux over the full data set and so may 
drop below the detection threshold.  Other transients  may not be detected  at all simply because 
the search is not optimized for the particularly short  timescale over which they were active. 

To search for transient/variable hard X-ray  sources in a systematic way,  Bird  et al. (2010, 2016) have specifically developed a 
method which optimizes their detection time-scale. It is the so-called bursticity  method where the \int/IBIS light curve for each source 
is scanned with a variable-sized  time window to search for the best source significance value. Then the duration and time interval, 
over which the source significance is maximized,  is recorded. 
As pointed out  by Bird et al. (2010, 2016),  the impact of this method has been very significant since it allowed 
to recover hundreds of transient sources which otherwise would have been missed  in the \int/IBIS surveys. 
They constitute an heterogeneous sample of transient/variable objects (most of which still unidentified),  especially in term of duration which ranges 
from very few hours to several months. The identification of these transient sources is much more 
difficult with respect to persistent ones as X-ray and optical/infrared  follow-ups  rarely provide a clear counterpart; 
this is because the source  very likely may be  in quiescence (and so undetectable) since the original \emph{INTEGRAL} detection. 
Despite these difficulties, such objects listed in the latest 1,000 orbits \int/IBIS catalog (Bird et al. 2016)  are very interesting because it is most probably among them that peculiar sources, or even a new class of objects, could emerge (e.g. Sguera et al. 2005, 2006). 

Here we report results from our study  focussed on the search for fast hard X-ray transients (i.e. duration less than a very few days) among the entire list of   transient/variable sources found with the bursticity method, as reported in the 1,000 orbits \int/IBIS catalog (Bird et al. 2016). 
From  such  list, we have extracted a  sample of  27 objects  satisfying the following  criteria: i) location on the Galactic plane at b$\leq$10$^{\circ}$, ii) duration of the outburst activity less than $\simeq$ 4 days, as recorded from the bursticity method, iii) unidentified nature, with unknown counterpart at lower energies. When available, we have used archival soft X-ray observations  (0.3--10 keV) which are useful to reduce the error circles to arcsecond sizes.  
This latter step is  mandatory to unequivocally individuate the infrared/optical counterpart of the source by using  archival data.  
All the collected information (e.g. temporal and spectral X-ray characteristics, duty cycle value, spectral type of the candidate infrared/optical counterpart) have been used to individuate, among the 27 objects,  those with  characteristics strongly resembling   Supergiant Fast X-ray Transients (SFXTs, Sguera et al. 2005, 2006, Negueruela et al. 2006), which are a newly discovered class of transient High Mass X-ray Binaries (HMXBs) showing a peculiar X-ray behavior  (for recent reviews see  Sidoli 2017,  Martinez-Nunez et al. 2017, Walter et al. 2015). Using this approach we were able to individuate four unidentified transients, from the original sample of 27 objects,  which can be considered good candidate SFXTs. All the pertaining results are reported in Section 3.

 \section{Observations and Data Reduction}
 
Throughout the paper, spectral analysis was performed using
XSPEC version 12.9.0 and, unless stated otherwise, the uncertainties are given at 90\% confidence 
for one interesting parameter. 

In order to find the lower energy counterparts of each transient source considered in this work,  we mainly used near infrared (NIR) archival information from surveys which nominally cover the Galactic plane:  UKIDSS GPS (Lucas et al. 2008),   VVV (Vista Variables in the Via Lactea, Minniti et al. 2010), 
GLIMPSE (Galactic Legacy Infrared Midplane Extraordinaire, Churchwell et al. 2009),  2MASS (Skrutskie et al. 2006).  
All such surveys have different depth, spatial and temporal coverage of the Galactic plane, hence each offers specific and complementary advantages in the search for the NIR counterparts. 
 
\subsection{INTEGRAL}
SFXT hunting is not an easy task because of their  very transitory nature and very low duty cycle, e.g. (0.1--5)$\%$ (Sidoli $\&$ Paizis 2018). 
Notably the IBIS/ISGRI instrument  on board \int~has repeatedly proven its suitableness for the discovery of new SFXTs.
Its instrumental characteristics (especially the good sensitivity on short timescales and the very large field of view, FoV) as well as its observational strategy (regular monitoring of the entire Galactic plane with many Ms of exposure time)
are  particularly suited in serendipitously detecting  short duration random sources such as SFXTs.

The temporal and spectral behaviour of each transient source investigated in this work has been studied in detail with  the ISGRI detector (Lebrun et al. 2003), which is the lower energy layer of the IBIS coded mask telescope (Ubertini et al. 2003) onboard \int~(Winkler et al. 2003, 2011).  In order to infer their  duty cycles, we have used the \textit{INTEGRAL} public-data archive described in Paizis et al (2013, 2016). 

 \emph{INTEGRAL} observations are divided into short pointings (Science Windows, ScWs) whose 
typical duration is $\sim$ 2,000 seconds.  Spectra and light curves were  extracted during the detected outbursts, the  data reduction was carried out with the release 10.2 of the Offline Scientific Analysis software (OSA, Courvoisier et al. 2003).  The IBIS/ISGRI systematics, which  are typically of the order of 1$\%$,  were added  to all extracted spectra. IBIS/ISGRI flux maps for each ScW during which the source was detected in outbursts were generated in the energy band that maximize the significance detection value, i.e. 18--60 keV.   Count rates at the position of
the source were extracted from individual ScW flux maps, to build light curves at ScW level.  
We applied a 12$^{\circ}$ limit because the response of IBIS/ISGRI is not well modelled at large off-axis
values and this may introduce a systematic error in the measurement of the source fluxes.  

Images from the X-ray Monitor JEM--X (Lund et al. 2003) were created for all the IBIS/ISGRI outbursts reported in this work, only in two cases they 
were inside  the JEM--X FoV. 

\subsection{\swift/XRT}

The Neil Gehrels Swift observatory (hereafter \swift, Gehrels et al. 2004) carries three instruments, one of which is the X-ray Telescope XRT (Burrows et al. 2005). We used archival \swift/XRT observations covering the position of the sources, they  were reprocessed with standard procedures using {\sc xrtpipeline} v0.13.4 included in the {\em HEASoft} software package version 6.25.
The low count rate of these sources allowed us to consider photon-counting data (PC) only.
We used the appropriate spectral redistribution matrices available in the Calibration
Database maintained by the High Energy Astrophysics Science Archive Research Center (HEASARC).
Source events were extracted from a circular region with a radius of 20 pixels,
local background events were taken within annular regions centred on the source 
(with inner and outer radii of 30 and 60 pixels respectively).
When fitting the X-ray spectra in {\sc xspec}, we adopted the photoelectric
absorption cross sections of  Verner et al. (1996) and the interstellar abundances of  Wilms et al. (2000), and
the model {\sc TBabs}.

The spectra were rebinned such that at least 20 counts per bin were present, 
to apply the $\chi^{2}$ 
statistics, and analysed in the energy range 0.3--10~keV.

The uncertainties on the source fluxes have been calculated  
fixing both low energy absorption and photon index to their most extreme values 
from their confidence contour levels (at 99\%), then
re-fitting the spectra to get the appropriate power law normalization.

\section{Results}
In the following we report results on the 4 unidentified hard X-ray transients selected with the methodology described at the end of section 1:
IGR~J16374$-$5043, IGR~J17375$-$3022, IGR~J12341$-$6143 and XTE~J1829$-$098.

\subsection{IGR~J16374$-$5043}
IGR~J16374$-$5043 was discovered with  \emph{INTEGRAL} in 2010 August during transient activity characterized by 
a short flare (Pavan et al. 2010).  A subsequent  \swift/XRT~follow-up, performed $\sim$ 1.5 days later, detected a fading soft X-ray counterpart (Bozzo et al. 2010).  Such reported brief communications (astronomer's telegrams) contained only very short information about IGR~J16374$-$5043, in particular the transient hard X-ray activity detected by \emph{INTEGRAL} was obtained from analysis of near real time data. On the basis of its variability,  IGR~J16374$-$5043 was selected by 
Cowperthwaite et al. (2013)  as member of a large sample of objects investigated for a possible gamma-ray blazar nature, but for IGR~J16374$-$5043 such investigation was unsuccessful.

Here we present the results of a more detailed  spectral and temporal analysis of the consolidated \int~data, together with a detailed investigation performed by using archival  \swift/XRT  as well as optical/infrared data. 

\subsubsection{INTEGRAL}
IGR~J16374$-$5043 is listed in the latest   \emph{INTEGRAL/IBIS} catalog (Bird et al. 2016) as a transient hard X-ray source found with the 
bursticity method. It is located on the Galactic plane at b$\sim$$-$2$^\circ$.5.  The outburst activity started on MJD 55430 or 22 August 2010 (satellite revolution 959), the source 
was best detected by  IBIS/ISGRI in the energy band 18--60 keV with a significance of $\sim$ 12$\sigma$ (3.5 ks effective exposure on-source). No detection was achieved in the higher energy band 60--100 keV. 

The IBIS/ISGRI light curve at ScW level (Fig. 1) clearly shows a fast X-ray flaring behaviour which reached at its peak a flux of 96$\pm$13 mCrab (18--60 keV) or 
$\sim$1.2$\times$10$^{-9}$ erg cm$^{-2}$ s$^{-1}$. The duration can be well constrained as $\sim$  12 hours.  The source was still in the IBIS/ISGRI FoV  before and after the temporal interval shown in the  light curve, however there was no sign of significant  flaring activity. 

We extracted the average IBIS/ISGRI spectrum during the outburst. 
The best fit is achieved with a power law model ($\Gamma$=2.1$\pm$0.4, $\chi^{2}_{\nu}$=0.9, 7 d.o.f.) 
The average 18--60 keV (20--40 keV) flux is $\sim$ 3.2$\times$10$^{-10}$ erg cm$^{-2}$ s$^{-1}$ (2$\times$10$^{-10}$ erg cm$^{-2}$ s$^{-1}$). 
Fig. 2  shows the power law data-to-model fit with the corresponding residuals.

IGR~J16374$-$5043 is not detected as a persistent source in the \emph{INTEGRAL/IBIS} catalog of Bird et al. (2016), despite extensive coverage of its sky region ($\sim$4.5 Ms). This information can be used to infer a 3$\sigma$ upper limit on its persistent hard X-ray emission, which  is 
of the order of 0.3 mCrab or 2.3$\times$10$^{-12}$ erg cm$^{-2}$ s$^{-1}$ (20--40 keV). When assuming the source peak flux as measured by IBIS/ISGRI 
from the outburst, we can derive a source dynamic range $\ge$ 520.

Following Paizis et al. (2013, 2016), we have searched the entire currently available IBIS/ISGRI public data archive from revolution 25 (Dec 2002) to 1919 (Feb 2018)  for possible additional short outbursts of IGR~J16374$-$5043  detected at ScW level. No significant detections have been obtained, which could have been associated with very short flares, above a significance value of 6$\sigma$ in different  energy bands (22--50, 18--50  and 50--100 keV).  The source exposure time  from the entire public archive is of the order of $\sim$ 10.5 Ms. The inferred duty cycle is  $\sim$ 0.4$\%$.

\begin{figure}
\begin{center}
\includegraphics[height=8.5 cm, angle=270]{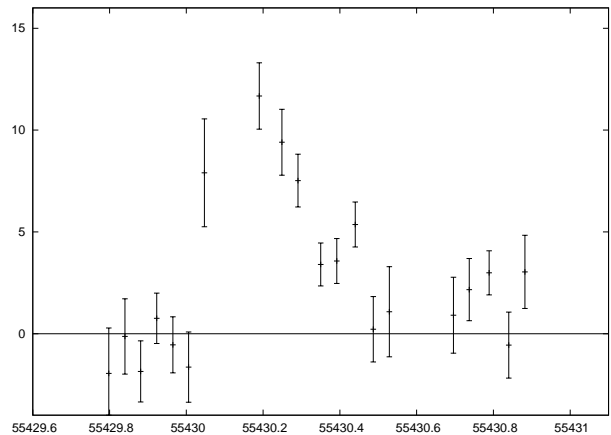}
\caption{IBIS/ISGRI light curve of IGR~J16374$-$5043 (18--60 keV) at ScW level (2,000 s) during  the detected outburst activity}
\end{center}
\end{figure}

\begin{figure}
\begin{center}
\includegraphics[height=8.5 cm, angle=270]{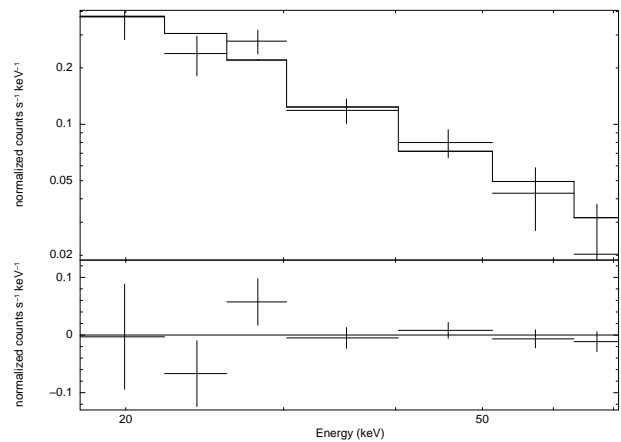}
\caption{IBIS/ISGRI spectrum of IGR~J16374$-$5043 extracted during the outburst, it is best  fitted by a power law model. The lower panel shows the residuals from the fit.}
\end{center}
\end{figure}

\subsubsection{{Archival Swift/XRT} and  infrared/optical data}

The \swift~satellite (Gehrels et al. 2004) performed a ToO observation of  IGR~J16374$-$5043  $\sim$ 1.5 days later its discovery with 
\int. The source was observed with the X-ray instrument  XRT  on 23 Aug  2010 at 16:43:00 UTC (150 s exposure, 3.2$\sigma$ detection, 0.3--10 keV) 
and 24 Aug 2010 at 00:29:00 UTC (1.8 ks exposure, 6.3$\sigma$ detection, 0.3--10 keV).  The exposure time of the first snapshot  (obs ID 31796001)  is too short to detect the source, so we focussed on the spectroscopy of the second
observation (obs ID 31796002), leading to a net count rate 
of (2.09$\pm{0.35}$)$\times10^{-2}$~counts~s$^{-1}$ (0.3--10 keV, but
note that below $\sim$1.4 keV there are no significant counts). 
Adopting Cash statistic in {\sc xspec}, after binning the spectrum at 1~count~bin$^{-1}$,
we fitted it  with an absorbed power-law model. We could place only an upper limit to the 
absorbing column density  (N$_{\rm H}<4\times10^{22}$~cm$^{-2}$),  while the 
photon index was equal to $0.76 ^{+1.70} _{-0.56}$ (C-stat = 31.03 for 30 d.o.f).
This model resulted into an observed flux equal to 3$\times10^{-12}$~erg~cm$^{-2}$~s$^{-1}$ (1--10 keV).
Such relatively low flux clearly indicates a rapid fading of the source observed after only $\sim$ 1.5 days after the outburst detected by \int.

\begin{figure}
\begin{center}
\includegraphics[height=7.2 cm]{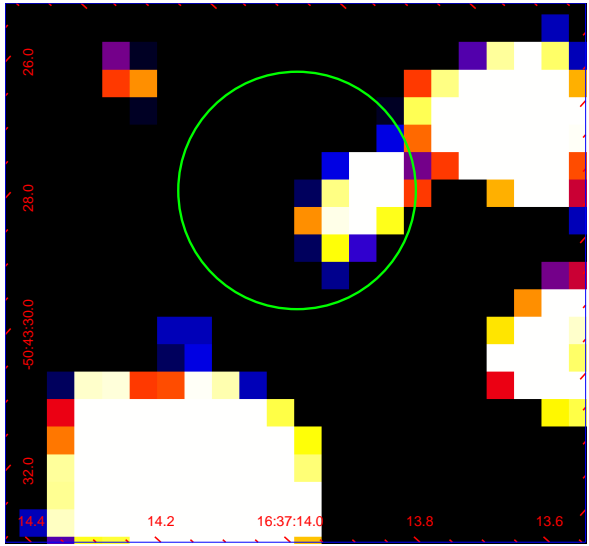}
\caption{\emph{Spitzer}  infrared map from the GLIMPSE  survey (3.6 $\mu$m band) of the sky region around IGR~J16374$-$5043, superimposed on  the \swift/XRT error circle at  90\%  confidence level}
\end{center}
\end{figure}

The best determined  \swift/XRT position is at R.A. (J2000)=$16^{\rm h}37^{\rm m}13\fs39$ 
Dec (J2000)=$-50^\circ 43\arcmin32\farcs6$ with a 90$\%$ confidence error circle radius of 2$''$.6 (using the XRT-UVOT alignment and matching UVOT field sources to the USNO-B1 catalog, see Evans et al. 2009).
Such refined position  allows us to perform a search for counterparts in the optical/infrared bands, by using all the available catalogues in the HEASARC data base. 
As a result,  we found that only one infrared source is located within the   XRT error circle (at 1$''$.5 from its centroid),  as  detected by \emph{Spitzer} 
during the mid-infrared (MIR) survey GLIMPSE (Galactic Legacy Infrared Midplane Extraordinaire). Such survey spans $\sim$ 130$^\circ$ in longitude (65$^\circ$ on either side of the center of the Galaxy) and $\sim$ 4$^\circ$  in latitude, providing magnitude measurements  in four different MIR wavelengths (3.6 $\mu$m, 4.5 $\mu$m, 5.8 $\mu$m, and 8 $\mu$m). The pinpointed best candidate MIR counterpart (source ID G334.8024-02.3838) has magnitude of 14.8 (3.6 $\mu$m) and 14.6 (4.5$\mu$m).  Fig. 3 shows the XRT error circle at  90$\%$  confidence  (2$''$.6 radius),   superimposed on 
the 3.6$\mu$m band \emph{Spitzer} image.  It is worth noting that  even if we consider the larger error circle at 99$\%$  confidence  (3$''$.7 radius), 
the MIR source G334.8024-02.3838 is still the only object located inside of it.  In the NIR band  (from 1.2$\mu$m to 2.1$\mu$m)  measurements from the surveys UKIDSS GPS and VVV are not available since the source position was not covered by their survey  area.  Conversely,  it was covered by the 2MASS all-sky survey with a spatial resolution comparable to that of \emph{Spitzer}.  The source is not listed in the 2MASS catalog  whose limiting magnitudes are 15.8, 15.1 and 14.3 in the  bands J, H, K, respectively. However we noted that the source is visible in the downloaded 2MASS JHK field images, although it must be weak with magnitudes 
below the catalog threshold. This, in conjunction with the \emph{Spitzer} detection,  likely suggests a very strong reddening. Unfortunately, obtaining useful magnitude estimates is not possible due to strong contamination from a bright nearby source.

 As for the optical band, we note that the MIR source G334.8024-02.3838 is listed in the \emph{GAIA}  catalog Data Release 2 (source ID 5940285090075838848) with  distance and magnitudes equal to  3.1$^{+2.8}_{-1.6}$ kpc, G=19.8,  G$_{BP}$=20.8 and 	G$_{RP}$=18.4, respectively. 
We note that \emph{GAIA} mean G passband covers a wavelength range from the near ultraviolet (roughly 330 nm) to the near infrared (roughly 1050 nm). The other two passbands, G$_{BP}$ and G$_{RP}$,  cover smaller wavelength ranges, from approximately 330 to 680 nm, and 630 to 1050 nm, respectively (Weiler 2018). If we assume the \emph{GAIA} optical source as counterpart of IGR~J16374$-$5043, then the 18--60 keV IBIS/ISGRI average (peak) luminosity   measured during the outburst is $\sim$ 4$\times$10$^{35}$ erg  s$^{-1}$ (1.5$\times$10$^{36}$ erg  s$^{-1}$). 
Conversely,  the  3$\sigma$ upper limit on the  persistent hard X-ray emission translates into a X-ray luminosity $\le$~2.7$\times$10$^{33}$ erg  s$^{-1}$.
Finally, the 1--10 keV luminosity during the \swift/XRT detection is equal to $\sim$ 3.4$\times$10$^{33}$ erg  s$^{-1}$.

\subsection{IGR~J17375$-$3022}

IGR~J17375$-$3022 is a transient hard X-ray source discovered with \int~in October 2008 during satellite revolution 732 (Ricci et al. 2008).
The source was not detected in the previous revolution as well as in the  following ones (Cadolle Bel et al. 2008, Ricci et al. 2008).  A  \swift/XRT follow-up was performed a few days later the \int~discovery, allowing to pinpoint a fading soft X-ray counterpart  (Beckmann et al. 2008).  
Notably, about six months later (in April 2009) the source was detected by chance with \emph{XMM-Newton}   while performing a slew between targets. 
The inferred 2--10 keV flux was of the order 
of $\sim$10$^{-10}$ erg cm$^{-2}$ s$^{-1}$ (Saxton et al. 2009).  
All the above information have been reported through brief communications,  in particular the transient hard X-ray activity detected by \emph{INTEGRAL} was obtained from analysis of near real time data. To date no detailed temporal/spectral analysis of the source 
has been reported in the literature, Here, we present the results of a  detailed  temporal/spectral analysis of the consolidated \emph{INTEGRAL} data, together with a detailed investigation of all available  archival  \swift/XRT   observation  (never published before) as well as  archival optical/infrared data.

\subsubsection{INTEGRAL}

\begin{figure}
\begin{center}
\includegraphics[height=8.8cm, angle=270]{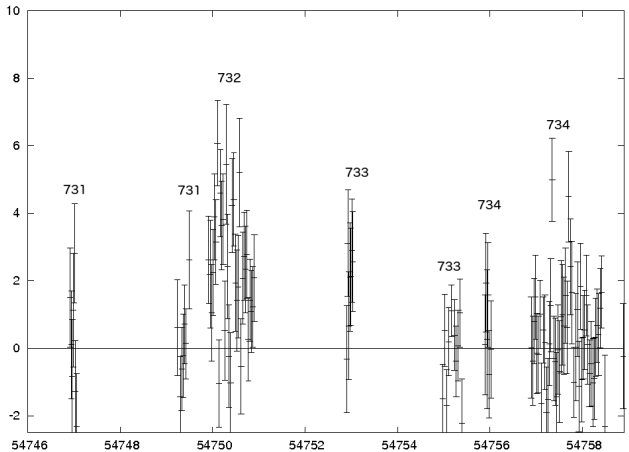}
\caption{IBIS/ISGRI light curve (18--60 keV) at ScW level ($\sim $ 2,000 s bin time) of  IGR~J17375$-$3022  covering the period from  revolution 731 to 734}
\end{center}
\end{figure}

IGR~J17375$-$3022 is listed in the latest IBIS/ISGRI catalog (Bird et al. 2016) as a transient hard X-ray source detected  with the 
bursticity method.  Its outburst activity started around the beginning of satellite revolution 732 (MJD 54749.9 or 10 October 2008 21:30 UTC), during which 
the most significance detection was obtained  in the band 18--60 keV (8.8$\sigma$, 18.5 ks effective exposure on-source). The average flux was measured as  9.7$\pm$1.1 mCrab or 1.2$\times$10$^{-10}$ erg cm $^{-2}$ s$^{-1}$ (18--60 keV). No significant detection was achieved   in the higher energy band 60--100 keV. 

The source was also in the JEM--X1 FoV for a total effective exposure of $\sim$ 12 ks, however it was not detected in both energy bands 3--10 
keV and 10--20 keV. The inferred 3$\sigma$  upper limit is of the order of $\sim$ 6 mCrab or 9$\times$10$^{-11}$ erg cm$^{-2}$ s$^{-1}$ (3--10 keV).

Fig. 4 shows the IBIS/ISGRI light curve at ScW level, covering observations from revolution 731 to 734. Hard X-ray activity is particularly evident during revolution 732, when the source was significantly detected as reported in the previous paragraph. It reached a peak flux of 27.4$\pm$5.7 mCrab or $\sim$ 3.6$\times$10$^{-10}$ erg cm$^{-2}$ s$^{-1}$ (18--60 keV). Conversely, in the mosaic significance images pertaining to revolutions immediately before (731,  8 ks on source) and after (733, 12.5 ks on source),  the source was not significantly detected; we inferred 18--60 keV 3$\sigma$ upper limits   of $\sim$
5 mCrab and  $\sim$ 4 mCrab, respectively. A deeper upper limit of  $\sim$ 2 mCrab  was obtained  by summing together consecutive revolutions 733 and 734 
(45 ks on source). Considering such results, we can firmly constrain the duration of the outburst activity in the range 1--3 days.

The  IBIS/ISGRI spectrum extracted during the outburst is best fitted  with a power law model ($\Gamma$=1.8$\pm$0.4, $\chi^{2}_{\nu}$=1.0, 5 d.o.f.). The average 18--60 keV (20--40 keV) flux is $\sim$ 1.2$\times$10$^{-10}$ erg cm$^{-2}$ s$^{-1}$ (7$\times$10$^{-11}$ erg cm$^{-2}$ s$^{-1}$).

IGR~J17375$-$3022 is  not detected as a persistent source in the  IBIS/ISGRI catalogue of Bird et al. (2016),  despite extensive coverage of its sky region ($\sim$11.6 Ms). We  inferred a 3$\sigma$ upper limit on its persistent hard X-ray emission equal to  0.2 mCrab or 1.5$\times$10$^{-12}$ erg cm $^{-2}$ s$^{-1}$ (20--40 keV). When assuming the source peak flux as measured by IBIS/ISGRI 
from the outburst, we can infer a  dynamic range $\ge$ 530. 

Following Paizis et al. (2013, 2016), we have searched the entire currently available IBIS/ISGRI public data archive from revolution 25 (Dec 2002) to 1919 (Feb 2018)  for possible additional short outbursts of IGR~J17375$-$3022 detected at ScW level. No detection has been obtained  (which could have been associated with very short flares) above a significance value of 6$\sigma$ in different  energy bands (22--50, 18--50  and 50--100 keV).  The source exposure time  from the entire public archive is of the order of $\sim$ 27.8 Ms. The inferred duty cycle is  $\sim$ 0.9$\%$

\subsubsection{Archival {Swift/XRT} and  infrared/optical data}

\begin{table}
\caption {Summary of  {\itshape Swift/XRT}  observations of IGR~J17375$-$3022}
\label{tab:main_outbursts} 
\begin{tabular}{ccccc}
\hline
\hline    
N.                     &     Obs ID                          &    start time                                  &    exp                   &  offset      \\
                  &                                          &    (UTC)                                       &    (ks)                & (arcmin)  \\
\hline    
1                    &   00043581001              &     2012-09-11    02:42:00             &    0.6                     &  8.0 \\ 
2                    &   00043580001                 &  2012-09-11    03:03:00                  &     0.5                   &  9.0      \\  
3                  &   00031278003                 &  2009-04-28 16:55:00              &       1.1                &  1.6   \\        
4                    &   00031278001              &     2008-10-14  13:58:00                 &    2                      &  0.4 \\ 
5                    &   00031278002              &     2008-10-14   17:26:00              &   8                    &  2.8 \\ 
\hline
\hline  
\end{tabular}
\end{table}

\begin{figure}
\includegraphics[height=8.0cm, angle=-90]{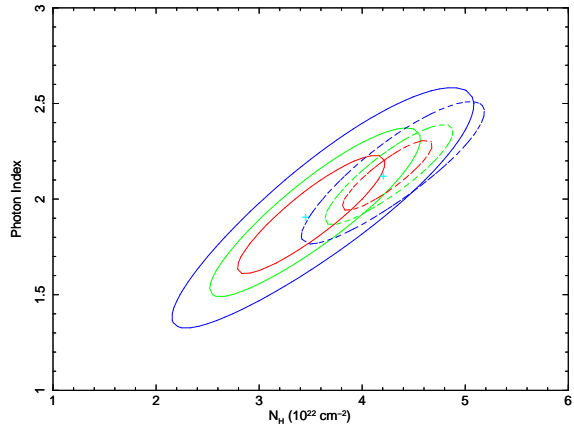}
\caption{Confidence contour levels (68\%, 90\% and 99\%) for the two parameters of the absorbed power law model
applied during the spectroscopy of IGR~J17375$-$3022 \swift/XRT observations ID 31278001 (solid lines) 
and 31278002 (dashed lines).
}
\label{igr17375_cont}
\end{figure}

The sky position of IGR~J17375$-$3022 was covered by the \swift~observations reported in Table~1.
Obs n. 5 is reported  by Beckmann et al. (2008) in a short communication as follow-up to the \int~source discovery. 
All the remaining observations were never reported in the literature. 

We note that the source is detected only during two observations, n. 4  (obs ID 31278001) and n. 5 (obs ID 31278002),  with a 0.3--10 keV count rate of
0.356$\pm{0.013}$ and 0.227$\pm{0.054}$ counts~s$^{-1}$, respectively.

The results obtained fitting the spectra with an absorbed power law model are 
the following. During the first observations we measured a low energy absorption
N$_{\rm H}$=(3.5$^{+0.8} _{-0.7}$)$\times10^{22}$~cm$^{-2}$ and 
a power law $\Gamma$=$1.91 ^{+0.35} _{-0.32}$ ($\chi^{2}_{\nu}$/dof=1.128/32). The observed flux, not corrected for the absorption, is equal to 3.2$\times10^{-11}$~erg~cm$^{-2}$~s$^{-1}$ 
in both energy bands 0.3--10 keV and 1--10 keV. 
Conversely, the unabsorbed fluxes are 7.3$\times10^{-11}$~erg~cm$^{-2}$~s$^{-1}$ (0.3--10 keV) and
5.1$\times10^{-11}$~erg~cm$^{-2}$~s$^{-1}$ (1--10 keV).

The spectral analysis of the second observation provided the following parameters for the power law model:
N$_{\rm H}$=(4.2$\pm$0.5)$\times$10$^{22}$ cm$^{-2}$ and  $\Gamma$=2.1$\pm$0.2 
($\chi^{2}_{\nu}$/dof=0.786/82) with an observed flux of 2.0$\times$10$^{-11}$ erg cm$^{-2}$ s$^{-1}$, 
again in both energy ranges. 
The estimated unabsorbed fluxes were 6$\times10^{-11}$~erg~cm$^{-2}$~s$^{-1}$ (0.3--10 keV) and
3.6$\times10^{-11}$~erg~cm$^{-2}$~s$^{-1}$ (1--10 keV).
In Fig.~\ref{igr17375_cont} we show the confidence contour levels 
of the absorbing column density and power law photon index in both observations.

From the observation n. 3 (obs ID~31278003), we estimated a Bayesian 95\% upper limit (Kraft et al. 1991) to the source count rate
of 0.0028~counts~s$^{-1}$.   We used it in WeBPIMMS in order to estimate  a 0.3--10 keV observed (unabsorbed)  upper limit flux of 
2.5$\times$10$^{-13}$~erg~cm$^{-2}$~s$^{-1}$ ($6.3\times$10$^{-13}$~erg~cm$^{-2}$~s$^{-1}$).
We assumed the same spectral model as from the previous \swift/XRT observations. When considering  the highest source flux, as measured by 
\swift/XRT in the observation n. 5, we can infer a dynamic range of $\ge$130. 
As for the remaining two observations (n. 1 and 2) with large off axis angle and very low exposure,  we estimate the following shallower 95\% Bayesian upper limits for the source count rate: $<$ 0.006 counts s$^{-1}$ (obs ID 00043580001) and   $<$ 0.005 counts s$^{-1}$ (obs ID 00043581001).

We calculated the enhanced source XRT position where the astrometry is derived using UVOT field sources (Evans et al. 2009)\footnote{http:/www.swift.ac.uk/user$\textunderscore$objects}:  R.A. (J2000)=$17^{\rm h}37^{\rm m}33\fs79$ 
Dec (J2000)=$-30^\circ 23\arcmin13\farcs8$ with a  90$\%$ confidence (99$\%$) error radius of 1$''$.4 (2$''$.0).
Such refined position allows us to perform a search for counterparts in the optical/infrared bands, by using all the available catalogues in the HEASARC data base. 

No catalogued  optical source is located within the  XRT error circle, according to the USNO catalog. 
In the optical V band, a  lower  limit of V$>$21 can be inferred  from the  survey limit as reported in Monet et al. (2003).  Conversely,  8 NIR sources are located inside the 99$\%$ confidence XRT error circle, as listed by the UKIDSS GPS Data Release 6. This survey, which started in 2005 May, covered  about  1,800 square degrees of the Galactic plane in the J, H, and K  filters (from 1 to 3 $\mu$m) with an angular resolution of 0.8 arcsec and down to a magnitude depth of $\sim$19,  i.e.  3 magnitudes deeper and more than a factor of 2 sharper than previous infrared surveys such as GLIMPSE or 2MASS. 
Importantly, the sub-arcsecond spatial resolution of UKIDSS is particularly suited  to identify the correct NIR candidate
counterpart to the hard X-ray source. All such objects  are  listed  in Table 2, while  Fig. 6  shows  the XRT error circles  superimposed on the  UKIDSS image ($K$ band).

\begin{figure}
\begin{center}
\includegraphics[height=7.6cm, angle=0]{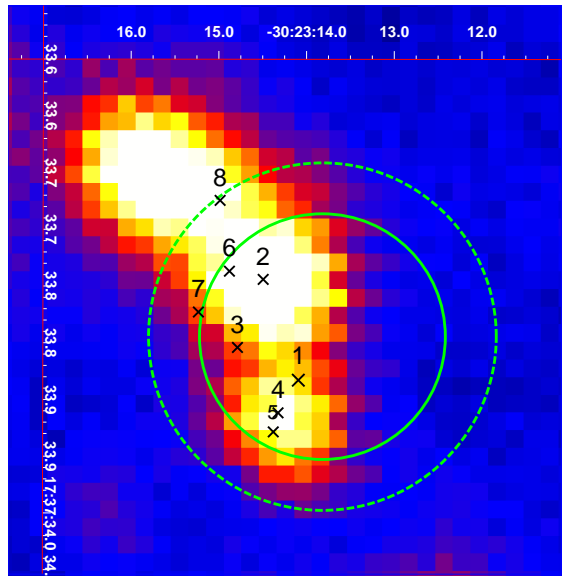}
\caption{UKIDSS  infrared map (K band) of the sky region around IGR~J17375$-$3022, superimposed on  the \swift/XRT error circles at  90\% and  99\% confidence level. Sources numbered from 1 to 8 are listed in Table 2}
\end{center}
\end{figure}

\begin{figure}
\begin{center}
\includegraphics[height=7.2cm, angle=0]{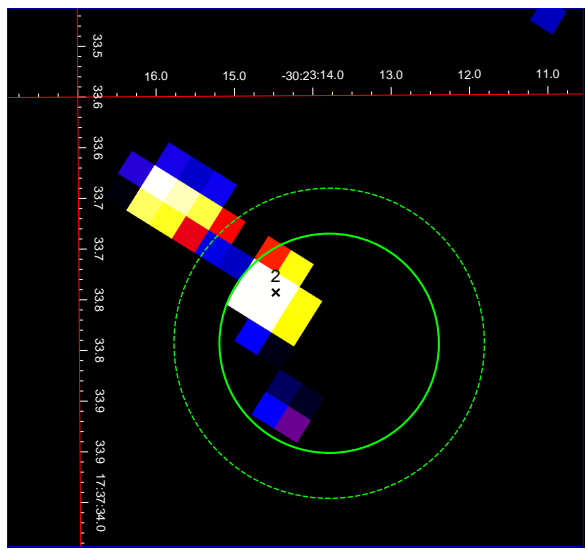}
\caption{VVV   infrared map (K band) of the sky region around IGR~J17375$-$3022, superimposed on  the \swift/XRT error circles at  90\% and  99\% confidence level}
\end{center}
\end{figure}

\begin{table*}
\caption {List of NIR sources (as taken from the UKIDSS GPS) located inside the 90\%  confidence  error circle (from n. 1 to n. 6)  and 99\%  confidence  error circle (from n. 7 to n. 8)  of IGR~J17375$-$3022. The table lists their  JHK magnitudes (lower limits  are derived according to Lawrence et al. 2007), offset from the XRT coordinates and Q value. $\dagger$  source also reported in the  VVV survey as   J173733.74-302314.55. 
}
\begin{tabular}{ccccccc}
\hline
\hline   
n.    & name                                 &    J                                 &   H                            &       K                                   & offset        & Q           \\
      &                                            &    (mag)                        &   (mag)                     &       (mag)                           &                     &                 \\
\hline    
1     &  J173733.83-302314.0	  &    22.144$\pm$3.645        &      $>$19.0             &     $>$18.8                      &   0.54$''$     &                 \\ 
2$\dagger$   &  J173733.74-302314.4      &    14.953$\pm$0.006   &   13.397$\pm$0.004   &   12.785$\pm$0.004       & 0.96$''$        &    0.09 
\\
3     &  J173733.80-302314.7     &    19.759$\pm$0.406    &    $>$19.0                  &     $>$18.8                        & 0.96$''$       &                \\
4     &  J173733.86-302314.3    &     16.190$\pm$0.017  &      $>$19.0                  &    13.922$\pm$0.012        &  1.02$''$      &                \\ 
5    &  J173733.87-302314.3    &      16.563$\pm$0.022   &    14.721$\pm$0.012  &     13.834$\pm$0.012       & 1.26$''$       &  0.16           \\
6     &  J173733.73-302314.8	 &    17.645$\pm$0.058   &    13.510$\pm$0.004  &   12.750$\pm$0.005         &   1.32$''$       &   2.69              \\ 
\hline    
7    &  J173733.77-302315.2    &   $>$19.9                     &   $>$19.0                  &   $>$18.8                       & 1.44$''$           &                \\
8     &  J173733.67-302314.9.  &    18.482$\pm$0.125   &   $>$19.0                 &     $>$18.8                   &   1.92$''$           &                  \\ 
\hline
\hline$\dagger$    &  	VVV J173733.74-302314.55     & 14.910$\pm$0.026         &       13.342$\pm$0.016         &         12.569$\pm$0.012      &   0.96$''$  & 0.09   \\ \hline  
\end{tabular}
\end{table*} 

In Table 2, we note that sources n. 1, 3, 4 (located inside the   90$\%$ confidence XRT error circle) and  7,  8  (located  outside the  90$\%$ confidence XRT error circle but inside the 99$\%$  one)  are particularly faint in the $J$  band and  even undetected in the $H$ and $K$ bands.  
Conversely, the remaining objects (n. 2,  5 and 6) are the brightest ones and  below we investigate the possibility that they could be (or not) the best candidate counterparts. \\

As for the brightest infrared object  detected by UKIDSS (n. 2 in table 2),  we note that it is  also reported in the NIR catalog  
VVV Data Release 2 (Minniti et al. 2017) as  J173733.74$-$302314.55.  The same does not hold for all the remaining infrared sources listed in table 2. The angular separation between the  two corresponding  coordinate centroids is only 0.$''$08. If we consider both the nominal positional accuracy of UKIDSS (0$''$.1, but it deteriorates to 0$''$.3 near the Galactic bulge, Lucas et al. 2008) and of VVV (0$''$.17)  then we can safely state that we are dealing with the same NIR source detected by both UKIDSS and VVV. The Vista Variables in the Via Lactea (VVV) is foremost a multi-epoch survey of the inner  regions  of  the  Milky  Way (bulge and adjacent sections of the Galactic plane) started in 2010 and spanning about five years. It was designed to complement other NIR single epochs survey (e.g. UKIDSS)  providing time domain information. Fig. 7 shows the VVV  infrared map (K band) of the sky region around IGR~J17375$-$3022 superimposed on  the 
\swift/XRT error circles at 90$\%$ and 99$\%$. It is evident that  J173733.74-302314.55 is the only detected NIR object inside of it.

If we calculate the reddening-free NIR quantity $Q$ (Negueruela \& Schurch 2007) with the extinction law of Fitzpatrick (1999), we note in table 2 that sources n. 5 and 6  have  a $Q$ value typical of intermediate or late type stars (Negueruela \& Schurch 2007). Conversely, the $Q$ value of source n. 2, calculated with the VVV magnitudes,  is typical of early-type stars OB . Following this latter indication, we can obtain a rough estimate of the distance of  source n. 2  (and its extinction, A$_{V}$) from its  J,H,K magnitudes as reported in table~2 and measured by VVV, assuming a stellar B spectral type and a specific luminosity class. 
In the hypothesis  of a supergiant  nature (for instance a B0.5Ib star with an absolute magnitude M$_V$=$-$6.8 mag, V-K=$-0.7$ and V-J=$-0.54$),
it should be located at a distance d$\sim$26~kpc (and with an A$_{V}\sim$14.5 mag) to match with the VVV near infrared magnitudes. 
On the other hand, a B0V star (in the hypothesis of a  Be HMXB nature, without taking into account the
possible contribution from the Be decretion disc) should be placed at $\sim$6~kpc distance (A$_{V}\sim$15 mag).
An intermediate distance of $\sim$10~kpc is allowed if we assume a giant nature for the B-type companion star.

In the light of the findings reported above, we propose the  infrared source n.2 (which  is the brightest one inside the X-ray error circle) as best candidate counterpart of IGR~J17375$-$3022 at lower frequencies, to date.  All fainter infrared objects reported in Table~2 should be located much farther away to reconcile with the observed magnitudes in these cases (e.g. making unlikely a supergiant donor). Additional X-ray observations  are  strongly needed (e.g. with 
\emph{Chandra}) in order to achieve a finer position which could  confirm or reject our  proposed counterpart.

\subsection{IGR~J12341$-$6143}
\subsubsection{INTEGRAL}

IGR~J12341$-$6143 is a very poorly studied hard X-ray transient. To date, no detailed information have been reported in the literature yet. The source has been discovered with \int~as listed in the latest IBIS catalog of Bird et al. (2016).  IGR~J12341$-$6143 was not detected as a persistent source despite extensive coverage of its sky region  ($\sim$3.5 Ms), the corresponding  3$\sigma$ upper limit is  equal to  0.2 mCrab or 1.5$\times$10$^{-12}$ erg cm$^{-2}$ s$^{-1}$ (20--40 keV). Conversely, IGR~J12341$-$6143 was detected with the bursticity method as a transient hard X-ray source. Its outburst activity started on MJD 54649.9 (2 July 2008 21:36 UTC) during  satellite revolution 698. The active source was in the IBIS/ISGRI FoV until MJD 54650.7, 
for an effective exposure time of 21 ks. It was detected at $\sim$ 6$\sigma$ level in the energy band 18--60 keV, with a measured flux of 5.6$\pm$0.9 mCrab.  No significant detection was achieved   in the higher energy band 60--100 keV. The obtained IBIS/ISGRI coordinates are RA=188$^\circ$.545  DEC=--61$^\circ$.708 with an error circle radius of 3$'$.98 at 90$\%$ confidence level.

Fig. 8 shows the 18--60 keV IBIS/ISGRI light curve at ScW level, covering observations from revolution 696 to 699. Hard X-ray activity is evident during revolution 698, when the source was discovered and significantly detected as reported in the previous paragraph. It reached a peak flux of 15.4$\pm$4.4 mCrab or $\sim$ 
2$\times$10$^{-10}$ erg cm$^{-2}$ s$^{-1}$ (18--60 keV).  When assuming the source upper limit on its persistent emission, we can infer a 
dynamic range $\ge$ 80. We note that the source was not significantly detected in the mosaic significance images pertaining to revolutions immediately  after (699, $\sim$ 17 ks effective exposure), and before (696, $\sim$ 16 ks effective exposure),  when it was again in the IBIS/ISGRI FoV. We derived a similar 18--60 keV 3$\sigma$ upper limit   of $\sim$  2 mCrab during each single revolution. Considering such findings, we can firmly constrain the duration of the outburst activity of IGR~J12341$-$6143 in the range 0.8--8 days.

The  IBIS/ISGRI spectrum extracted during the outburst is best fitted  with a thermal bremsstrahlung model 
with temperature kT=14$^{+19}_{-7}$ keV ($\chi^{2}_{\nu}$=1.02, 3 d.o.f.). The average 18--60 keV (20--40 keV) flux is equal to $\sim$ 5.2$\times$10$^{-11}$ erg 
cm$^{-2}$ s$^{-1}$ (3.6$\times$10$^{-11}$ erg cm$^{-2}$ s$^{-1}$).  Fig. 9  shows the  bremsstrahlung data-to-model fit with the corresponding residuals.
Alternatively, a power law model provides a reasonable description ($\chi^{2}_{\nu}$=1.2, 3 d.o.f.) with a soft photon index not well constrained 
($\Gamma$=3.7$^{+2.0}_{-1.3}$). 

\begin{figure}
\begin{center}
\includegraphics[height=8.8cm, angle=270]{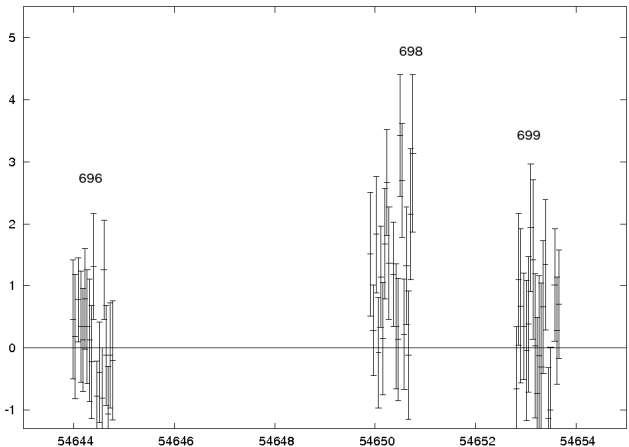}
\caption{IBIS/ISGRI light curve (18--60 keV) at ScW level of  IGR~J12341$-$6143  covering the period from  revolution 696 to 699}
\end{center}
\end{figure}
\begin{figure}
\begin{center}
\includegraphics[height=8cm, angle=270]{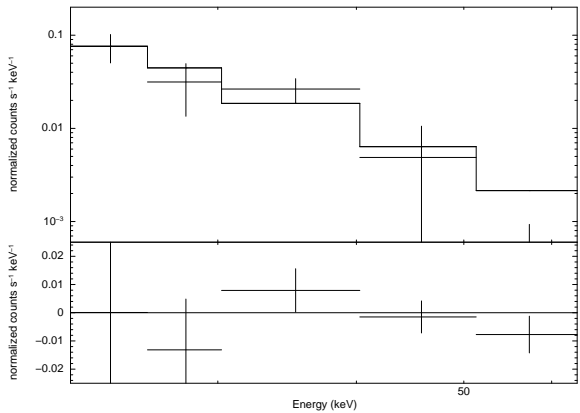}
\caption{IBIS/ISGRI spectrum of  IGR~J12341$-$6143 extracted during the outburst, it is best  fitted by a bremsstrahlung model. The lower panel shows the residuals from the fit}
\end{center}
\end{figure}

IGR~J12341$-$6143 was also in the JEM--X1 FoV during the detected outburst activity in revolution 698, for an effective exposure of $\sim$ 17 ks. 
The source was barely detected at $\sim$ 4$\sigma$ level in the energy band  3--10 keV, no significant detection was achieved in the higher energy band 10--20 keV. The measured 3--10 keV flux is equal to 3.7$\pm$0.9 mCrab or  $\sim$ 5.5$\times$10$^{-11}$ erg cm$^{-2}$ s$^{-1}$.  Unfortunately the source was too faint
and  the very low  statistics prevented the extraction of a meaningful spectrum and a proper study of the source. 

Following Paizis et al. (2013, 2016), we have searched the entire currently available IBIS/ISGRI public data archive from revolution 25 (Dec 2002) to 1919 (Feb 2018)  for possible additional short outbursts of IGR~J12341$-$6143 detected at ScW level. No detection has been obtained  (which could have been associated with very short flares) above a significance value of 6$\sigma$ in different  energy bands (22--50, 18--50  and 50--100 keV).  The source exposure time  from the entire archive is of the order of $\sim$ 7.7 Ms. The inferred duty cycle is  $\sim$ 0.9$\%$

\subsubsection{Targeted \swift/XRT observation}

Before our work, the field position of IGR~J12341$-$6143 has never been covered  below 10 keV by any soft X-ray mission. 
Hence we asked for a ToO observation of the sky region with the \swift~satellite, with the aim of eventually obtaining for the first time useful information from the soft X-ray band.  The \swift/XRT ToO observation was performed on 4 Sep 2019 at 23:11:34 UTC with an exposure  of $\sim$ 1 ks. The source was not detected, nonetheless the observation was  useful to estimate  the  Bayesian 95\%  upper limit (Kraft et al. (1991) to the source count rate equal to   
0.005  counts/s.  Using WebPIMMS, it converts into a flux of  4.6$\times$10$^{-13}$ erg cm$^{-2}$ s$^{-1}$
by adopting the average absorbing column density along the line of sight (Dickey \& Lockman1990) and 
a power-law photon index of 1.  When assuming the outburst source flux, as measured by \int/JEM--X in a similar energy band, 
we can infer a dynamic range $\ge$ 120. 

To date, the only usable source position is that provided by IBIS/ISGRI with a large error circle radius of $\sim$ 4$'$. Since IGR~J12341$-$6143 
 is located on the Galactic plane in a very crowded sky field, the lack of an accurate arcsecond  sized localization prevents us from pinpointing the correct unique NIR/optical counterpart among the several tens candidates.

\subsection{XTE~J1829$-$098}

XTE~J1829$-$098 is a transient X-ray pulsar (7.8 s) discovered in outburst by \emph{RXTE}/PCA  on July 2004 during scans of the Galactic plane (Markwardt et al. 2004). The 2--10 keV flux was of the order of 1.0$\times$10$^{-10}$ erg  cm$^{-2}$ s$^{-1}$, the X-ray spectrum was hard and absorbed. From regular \emph{RXTE}/PCA monitoring observations of the source, Markwardt et al. (2009) estimated a typical outbursts duration of $\sim$ 7 days. The source has been serendipitously observed by both  \emph{XMM}-Newton and \emph{Chandra} in several occasions (Halpern et al. 2007), providing measurements of   both  deep upper limit (2.5$\times$10$^{-14}$ erg  cm$^{-2}$ s$^{-1}$) and moderate X-ray fluxes ($\sim$ 5$\times$10$^{-11}$ erg  cm$^{-2}$ s$^{-1}$).  Notably, the inferred dynamic range of the source is particularly high, i.e. $\sim$ 6,800. The accurate \emph{Chandra} localization allowed the identification of  the likely infrared counterpart whose spectral type and luminosity class are still unknown although its characteristics seems to be  typical of an highly reddened O or B star (Halpern et al. 2007).  At the beginning of  August  2018 MAXI/GSC detected renewed X-ray activity from the source which reached a 4--10 keV flux of 24$\pm$4 mCrab (Nakajima et al. 2018). Such detection triggered a ToO observation with \emph{NuSTAR} (16 Aug 2018) which measured  an average 3--79 keV  flux of $\sim$ 3$\times$10$^{-10}$ erg  cm$^{-2}$ s$^{-1}$ (Shtykovsky et al. 2018). In particular, in the source spectrum a cyclotron absorption line was detected at E$_{cyc}$ $\sim$ 15 keV. All the above  findings clearly indicate a firm HMXB nature for the source. 

Here we present the \int~results of a detailed spectral and temporal analysis of the source in outburst, which is the first ever  study temporally covering an entire outburst above 20 keV. In addition, we performed an investigation of all available \swift/XRT archival observations (not published yet) as well as  archival infrared data. 

\subsubsection{INTEGRAL}

\begin{figure}
\begin{center}
\includegraphics[height=8.5cm, angle=270]{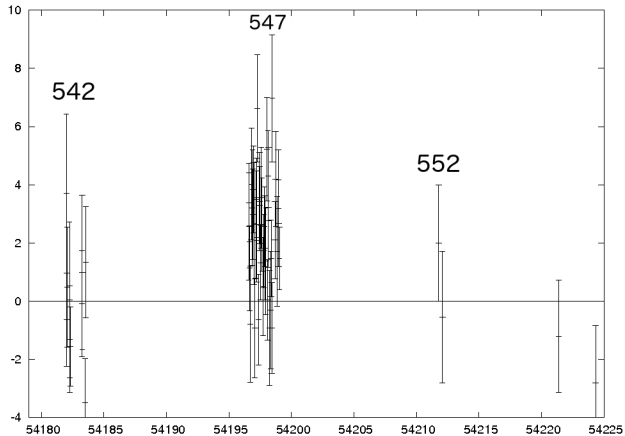}
\caption{IBIS/ISGRI light curve (18--60 keV) at ScW level of   XTE~J1829$-$098  covering the period from  revolution 542 to 552}
\end{center}
\end{figure}

XTE~J1829$-$098  is reported  in the latest   \emph{INTEGRAL/IBIS}  catalog  by Bird et al. (2016)  as a transient hard X-ray source detected  with the 
bursticity method. Its outburst activity started around MJD 54195 (or 5 April 2007) during satellite revolution 547. The source was detected at 10.4$\sigma$ level 
(18--60 KeV,  $\sim$ 19 ks effective exposure on-source), no significant detection was achieved   in the higher energy band 60--100 keV. 
The average flux was 10.80$\pm$1.04 mCrab or $\sim$ 1.4$\times$10$^{-10}$ erg cm$^{-2}$ s$^{-1}$ (18--60 keV). 

Fig. 10 shows the 18--60 keV  IBIS/ISGRI light curve at ScW level, covering observations from revolution 542 to 552.  Hard X-ray activity is evident during revolution 547, when the source was significantly detected as reported in the previous paragraph. It reached a peak flux of 31.4$\pm$9.8 mCrab or $\sim$ 4.1$\times$10$^{-10}$ erg cm$^{-2}$ s$^{-1}$ (18--60 keV).  Unfortunately the \emph{INTEGRAL }temporal coverage of the outburst is not particularly good, in fact  there are gaps in the light curve since for several days the source was not in the IBIS/ISGRI FoV before and after revolution 547. When the source was again in the  FoV (revolutions 542 and 552) it seems that it was not active anymore, although this cannot be firmly claimed because  only  few ScWs are  available.  Based on the \emph{INTEGRAL} data coverage,  the  duration of the outburst activity can be  loosely constrained  in the range 3--25 days. In this context, \emph{Swift/BAT} data can be helpful since the BAT instrument provides a continuos  time coverage of the source activity with no temporal gaps. 
From the \swift/BAT archive,  we downloaded the 15--50 keV source light curve  on daily  time-scale, and we checked if the source was active or not 
during the temporal range corresponding to the gaps of the IBIS/ISGRI light curve. No significant hard X-ray activity was evident, suggesting that the duration of the outburst  was  very likely of the order of 3--4 days. This  is roughly the same order of duration typically estimated by Markwardt et al. (2009) from RXTE/PCA monitoring observations ($\sim$ 7 days). 

The  IBIS/ISGRI spectrum extracted during the outburst is well fitted  with a power law model ($\Gamma$=3.3$\pm$0.6, $\chi^{2}_{\nu}$=1.1, 3 d.o.f.). 
The average 18--60 keV flux is $\sim$ 9.2$\times$10$^{-11}$ erg cm$^{-2}$ s$^{-1}$.
Alternatively,  a  thermal bremsstrahlung model provides a reasonable fit  with   kT=15$^{+7}_{-4}$ keV ($\chi^{2}_{\nu}$=0.6, 3 d.o.f.).

XTE~J1829$-$098  is  not detected as a persistent source in the  \emph{INTEGRAL/IBIS} catalogue of Bird et al. (2016),  despite extensive coverage of its sky region ($\sim$ 4.8 Ms). We  inferred a 3$\sigma$ upper limit on its persistent hard X-ray emission equal to  0.3 mCrab or 3.8$\times$10$^{-12}$ erg cm $^{-2}$ s$^{-1}$ (20--40 keV). When assuming the source peak flux as measured by IBIS/ISGRI  from the outburst, we can infer a source dynamic range $\ge$ 110. 

Following Paizis et al. (2013, 2016), we have searched the entire currently available IBIS/ISGRI public data archive from revolution 25 (Dec 2002) to 1919 (Feb 2018) for possible additional short outbursts of  XTE~J1829$-$098  detected at ScW level. No additional detections have been obtained above a significance value of 6$\sigma$ at ScW level in different  energy bands (22--50, 18--50  and 50--100 keV).  The source exposure time  from the entire archive is of the order of $\sim$ 8.2 Ms. The inferred duty cycle is  $\sim$ 4.2$\%$

\subsubsection{{Archival Swift/XRT observations}}

\begin{table}
\caption {Summary of  {\itshape Swift/XRT}  observations of XTE~J1829$-$098}
\label{tab:main_outbursts} 
\begin{tabular}{ccccc} 
\hline
\hline    
N.                     &     Obs ID                          &    start time                                  &    exp                   &  offset      \\
                  &                                          &   (UTC)                                       &    (ks)                & (arcmin)  \\
\hline    
1                    &   00087499003            &     2019-03-21  12:27:00                 &    1                      & 8.8 \\ 
2                    &   00087498002                 &  2018-02-21    05:03:00           &     0.7                   &  7.2      \\  
3                    &   00010806001             &     2018-08-24    09:10:00             &    0.5                     & 2.6 \\ 
4                     &  00087498001             &     2017-10-07    12:27:00             &    4.2                     & 8.9 \\ 
5                    &   00034147001              &     2015-11-05   18:15:00              &   0.5                    &  9.8 \\     
6                  &     00044284001              &     2012-11-11 14:20:00              &       0.5                &  8.2   \\     
7                    &   00044277001              &     2011-05-16   03:27:00              &   0.5                    &  9.3 \\

\hline
\hline  
\end{tabular}
\end{table}

\begin{table*}
\caption {Summary of characteristics related to the studied X-ray transients and to their outbursts}
\begin{tabular}{ccccccc}
\hline
\hline   
name                              &     duration     &         outburst average flux                  &  ouburst average luminosity                      &  duty           & dynamic & dynamic    \\
                                       &    (days)       &                             (18--60 keV)                  &  (18--60 keV)                 &  cycle      &   range    &    range  \\
                                       &                    &        (erg cm$^{-2}$ s$^{-1}$)                    & (erg s$^{-1}$)                 &                &   ($>$ 18 keV)          &    (0.3--10 keV)     \\
\hline    
IGR~J16374$-$5043    &     0.5            &             3.2$\times$10$^{-10}$              &    4$\times$10$^{35}$    & 0.4$\%$   &   $>$522 &    \\ 
IGR~J17375$-$3022    &     1--3           &             1.2$\times$10$^{-10}$             &     9.7$\times$10$^{36}$      & 0.9$\%$   &   $>$533 &   $>$130 \\ 
IGR~J12341$-$6143     &    0.8--8       &              5.2$\times$10$^{-11}$          &                                              &    0.9$\%$   &   $>$77 &  $>$120  \\
XTE~J1829$-$098        &    4               &             9.2$\times$10$^{-11}$             &      3.5$\times$10$^{36}$    &  4.2$\%$     &   $>$110 &    6,800 \\
\hline    
\hline  
\end{tabular}
\end{table*} 

The sky position of XTE~J1829-098 fell within the \swift/XRT FoV several times, as reported in Table~3. To date  none of such observations has been reported in the literature. 

We note that only during one observation (n. 3, obs ID 00010806001)  was the source detected and a meaningful spectroscopy  possible. 
Given the very short exposure (T$_{exp}$=534~s) and the faint source intensity 
(net count rate 0.314$\pm{0.025}$~counts~s$^{-1}$), we adopted Cash statistics
when fitting the spectrum with an absorbed power law model. 
Also in this case, no significant net counts were observed below $\sim$1~keV with \swift.
We obtained an absorbing column density 
N$_{\rm H}$=(10$^{+6} _{-4}$)$\times10^{22}$~cm$^{-2}$, a photon index $\Gamma$=$1.1 ^{+0.9} _{-0.8}$
(C-stat = 171.80 for 421 dof), 
resulting in an observed flux of 4.8$\times10^{-11}$~erg~cm$^{-2}$~s$^{-1}$ (in both 0.3--10 keV and 1--10 keV bands).
The unabsorbed fluxes are 
8.7$\times10^{-11}$~erg~cm$^{-2}$~s$^{-1}$ (0.3--10 keV) and
7.9$\times10^{-11}$~erg~cm$^{-2}$~s$^{-1}$ (1--10 keV).

In all the remaining observations listed in Table 3, the source was not detected. However we note that in all but one observation the exposure time 
was particularly low (i.e. 0.5--1 ks). We estimated the Bayesian 95\%  upper limit (Kraft et al. (1991) to the source count rate 
from the observation with the longer exposure (4.2 ks, n. 4, obs ID 00087498001).  It was equal to 0.0025 count/s. 
Using WebPIMMS, it converts into an observed (unabsorbed)  0.3--10 keV   flux of  $\sim$ 4$\times10^{-13}$~erg~cm$^{-2}$~s$^{-1}$
($\sim$ 7.6$\times10^{-13}$~erg~cm$^{-2}$~s$^{-1}$). We assumed the same spectral model as from the  \swift/XRT observation n. 3.

\subsubsection{Investigation at lower frequencies with archival data}
Halpern et al. (2007), using the  precise \emph{Chandra} localization of XTE~J1829$-$098,  individuated its probable infrared counterpart through  images in the K and H bands as obtained with  the MDM 2.4 m telescope on 17 Jun 2005. The measured magnitudes were H$\sim$13.9 and K$\sim$12.7, no measurement was obtained in the J band. Back in 2005, no additional infrared measurements were available from other infrared surveys,  e.g. the UKIDSS survey started on 2005. Currently, we note that the infrared counterpart  proposed by Halpern et al. (2007) has been also detected during the  UKIDSS GPS data release 6  with magnitudes J=16.147$\pm$0.009,  H=14.064$\pm$0.003, K=12.591$\pm$0.002.  If we use the reddening-free NIR diagnostic $Q$ (Negueruela \& Schurch 2007) with the extinction law of Fitzpatrick (1999), we note that it has a  $Q$ value  of --0.7 which is very typical of early-type OB stars. In particular, the negative value  strongly indicates  an infrared excess. 

Adopting the UKIDSS infrared magnitudes of such proposed counterpart, we can
obtain a rough idea about its distance: if we assume a supergiant B0.5Ib star (with M$_V$=$-$6.8 mag, V-K=$-0.7$ and V-J=$-0.54$),
the source should be located at a distance d$\sim$18~kpc (with A$_{V}\sim$21 mag). 
This implies a V magnitude fainter than V$\sim$30--31 mag. A main sequence B0 star (assuming M$_V$=$-$4 mag, V-K=$-0.9$ and V-J=$-0.7$), would imply
a distance d$\sim$4.5~kpc (with A$_{V}\sim$22 mag), while a giant B0 star (assuming M$_V$=$-$5 mag, V-K=$-0.8$ and V-J=$-0.7$),
would be located at about 8~kpc.

\subsection{Search for periodicities}
We performed a timing analysis using  the  Lomb--Scargle   method  (Lomb  1976;  Scargle  1982) of the long term 
\int-IBIS and \swift-BAT light curves of the sources  under study, in order to search for periodicities which could eventually be ascribed to their orbital periods. 

For all four sources, the long term IBIS/ISGRI ScW  light curves (18--60 keV)  cover the observational period from 2003 March  to 2015 January. 
We applied standard optimum filtering in order to exclude poor quality data points which could eventually disrupt the periodic signal 
(see Goossens et al. 2013 and Sguera et al. 2007 for details). Periodicities were searched for in the range 1--100 days, but none were found. 

The public access  \swift/BAT light curves (15--50 keV) on daily timescale of all sources  but IGR~J12341$-$6143  were also searched for periodic signals in 
the range 2--250 days, but none were found. 

\section{Discussion}

Table 4  provides a  summary of  the X-ray characteristics of the four transients.  Although all the  reported information  do not allow a firm identification of  their nature, in the following we use them to obtain  hints and/or indications on the most likely class of X-ray sources to which they  could belong to. 

IGR~J16374$-$5043 has been detected only once (when discovered) during a bright (peak-flux of $\sim$10$^{-9}$ erg cm$^{-2}$ s$^{-1}$) and short ($\sim$12 hours) X-ray flare, strongly resembling those from SFXTs.  The low duty cycle and high dynamic range above 18 keV, the hard X-ray spectral shape in the soft X-ray band, all  are very typical of SFXTs as well (Sidoli $\&$ Paizis 2018, Sguera et al. 2008). We individuated a single optical/infrared source inside the arcsecond sized \swift/XRT error circle, as detected by \emph{Spitzer} and \emph{GAIA}, which is the  likely the  counterpart at lower frequencies. \emph{GAIA} measured a distance of $\sim$ 3.1 kpc, the corresponding average X-ray luminosities during the outburst (18--60 keV, 4$\times$10$^{35}$ erg s$^{-1}$) and during the rapid fading phase (1--10 keV, 3.4$\times$10$^{33}$ erg s$^{-1}$) are typical of SFXTs. The infrared counterpart is bright  in the MIR (as detected by Spitzer)  while it is weaker in the NIR (e.g. 2MASS), suggesting that it is affected by high intrinsic extinction. In the light of  the above findings, IGR~J16374$-$5043 can be considered a strong SFXT candidate.  NIR spectroscopy of the proposed counterpart is needed to definitely confirm this suggested nature.

IGR~J17375$-$3022 is a hard X-ray transient whose detected bright outburst had a duration constrained by \int~observations in the range 1--3 days, i.e.  similar to  a few  firm SFXTs (e.g. AX J1949.8+2534, Sguera et al. 2017; IGR J17354$-$3255, Sguera et al. 2011). Furthermore the low duty cycle, high dynamic range and spectral shape of the source,  all are very typical of SFXTs as well. We individuated an early type OB objects inside the arcsecond sized 
\swift/XRT error circle, as detected by UKIDSS/VVV surveys in the infrared,  which we proposed as the best candidate counterpart at lower frequencies, to date. 
In the hypothesis of a supergiant nature for this donor star, we found that it should be located at $\sim$ 26 kpc: the corresponding luminosities in outburst  (18--60 keV, 9.7$\times$10$^{36}$ erg s$^{-1}$; 0.3--10 keV, 2.6$\times$10$^{36}$ erg s$^{-1}$) and during the lowest X-ray state (0.3--10 keV, $<$2$\times$10$^{34}$ erg s$^{-1}$) are typical of SFXTs. Such large distance would place the source in the outer arm of the Galaxy, much like a few other cases of distant SFXTs, e.g. IGR~J18462$-$0223 (Sguera et al. 2013) and IGR~J16418$-$4532 (Drave et al. 2013).  It is likely that, due to the relatively large distance of the source, only the brightest outbursts with L$_{X}$$\ge$10$^{36}$ erg s$^{-1}$ (like the one reported in this study) are detectable by \int~while the lower intensity  ones (L$_{X}$$\sim$10$^{35-36}$ erg s$^{-1}$)  could be  too faint to be detected at energies above 18 keV. Overall, IGR~J17375$-$3022 can be considered a good candidate SFXT. Additional observations  are  strongly needed: i)  in the X-ray band in order to achieve a  sub-arcsecond sized position  which would allow to pinpoint a unique infrared counterpart, ii) in the infrared band in order to unveil the nature of the proposed best candidate counterpart through spectroscopy.

IGR~J12341$-$6143 has been discovered by \int~during an outburst whose duration is  constrained in the range 0.8--8 days.  Such  interval is compatible with both  SFXTs  (i.e. lower part of the interval)  or alternatively  Be HMXBs  (i.e. higher part of the interval). The spectral shape above 18 keV is very similar to HMXBs in general, in particular the low duty cycle is more typical of SFXTs rather than Be HMXBs. Unfortunately useful information below 10 keV are not available since the only observation performed to date (and reported in our work) did not provide any detection.  Consecutively, the only usable source position is that provided by IBIS/ISGRI with an error circle radius of $\sim$ 4$'$, i.e. way to much large to pinpoint a unique NIR/optical counterpart. In the light of the findings above, the most probable nature for the source is a HMXB, both  SFXT and Be HMXB are a viable scenario. 

The X-ray properties of the source XTE~J1829$-$098, especially its pulse period and the cyclotron line, 
indicate a firm HMXB nature.  Discovered in 2004, the source is characterized by a remarkable dynamic range of $\sim$ 6,800 below 10 keV. A typical duration of $\sim$ 7 days has been estimated for its outbursts through \emph{RXTE} monitoring.  To date only one detection  above 20 keV is reported in the literature, as obtained during a short  \emph{NuSTAR}  ToO observation in 2018 in response to a MAXI trigger of outburst activity. From archival \int~data, we reported on the second detection ever above 20 keV obtained during an outburst in April 2007, the first ever outburst studied  at hard X-rays during  its entire duration. The latter was of the order of a few days ($\sim$ 4),  constraining the typical historical duration of outbursts from the source in the range $\sim$ 4--7 days. This is apparently at odds with the  typical duration of classical outbursts from SFXTs (i.e. less than a day). However  a few other SFXTs are known to show unusually longer hard X-ray activity exceptionally lasting several days, e.g. IGR J18483$-$0311 (Sguera et al. 2015),  AX~J1949.8+2534 (Sguera et al. 2017), IGR~J17354$-$3255 (Sguera et al. 2011), IGR J17503$-$2636 (Ferrigno et al. 2019), i.e. comparable to the duration of the hard X-ray activity usually detected from  XTE~J1829$-$098.  
In addition, we note that the dynamic range of the source (very high in the soft X-ray band),  its duty cycle,  its spectral shape,  all are consistent with a SFXT nature as well. Inside the sub-arcsecond sized \emph{Chandra} error circle there is a unique and bright infrared counterpart  (Halpern et al. 2007). 
We found that it is very likely an early type OB star. In the hypothesis of a supergiant nature, it should be located at $\sim$ 18 kpc: we found that the corresponding source luminosities in outburst  (18--60 keV, 3.5$\times$10$^{36}$ erg s$^{-1}$; 0.3--10 keV, 6.6$\times$10$^{36}$ erg s$^{-1}$) and during the lowest X-ray state (0.3--10 keV, $\sim$1.0$\times$10$^{33}$ erg s$^{-1}$) are very typical of SFXTs.  Overall,  XTE~J1829$-$098 can be considered a good candidate SFXT.
Alternatively, we point out that in principle the hypothesis  of a Be HMXB nature is viable as well. In this case,  the infrared counterpart  would be located at   $\sim$ 4.5 kpc.  
However, in this scenario the corresponding outburst X-ray luminosities ($\sim$10$^{35}$ erg s$^{-1}$) and durations (4--7 days)  are about one order of magnitude lower that typical values of classical Be HMXBs.

\section{Summary and conclusions}

We performed a systematic study of the entire sample of   unidentified fast hard X-ray transients located on the Galactic plane and detected with the bursticity method, as reported in Bird et al. (2016). Our main aim was to  individuate those with  X-ray characteristics strongly resembling  Supergiant Fast X-ray Transients.  

We~found~that~IGR~J16374$-$5043 and IGR~J17375$-$3022, which before our current work were  very poorly studied in the literature, now can  be considered strong candidate SFXTs.   IGR~J12341$-$6143 was also very poorly studied in the literature before our  work.  We found that its  characteristics (in particular the outburst duration) are compatible with both  SFXT and Be HMXB scenarios.
Finally, before our current work the transient HMXB XTE~J1829$-$098   was studied in detail below 10 keV and  conversely  it was poorly investigated above 18 keV.  Our study above 18 keV, in conjunction with archival infrared data, allows  for the first time to consider  it  as  good candidate 
SFXT, although a Be HMXB nature cannot be entirely excluded. 

To date, about  14 firm SFXTs have been reported in the literature  along with a smaller number of candidates ($\sim$ 4). 
One of the main aims of the current studies on SFXTs is to increase the sample of firm objects, this is mandatory for population studies e.g. to establish  whether SFXTs are a homogeneous class or display a variety of different X-ray characteristics.  In this context, the findings of our work  significantly increase
 the sample of candidate SFXTs, effectively doubling their number.  NIR  spectroscopy of the proposed candidate counterparts is strongly  needed in order to firmly  
confirm (or reject) our suggested nature of SFXTs. It seems plausible that other SFXTs  wait to be discovered in our Galaxy. Further exploitations of the 
entire \int~data archive, which in the meanwhile doubled its dataset since the latest published \emph{INTEGRAL}/IBIS  catalog, may yield additional discoveries of this kind of interesting and peculiar X-ray transients.

\section*{Acknowledgments}
We thank the anonymous referee for a prompt report which helped us to improve the quality of this work.
VS thanks N. Masetti for useful discussions. 
We thank the Swift team, the PI, the duty scientists and science planners for making the ToO observation reported here possible.
We acknowledge financial support from the Italian Space Agency via ASI--INTEGRAL agreement n. 2013-025.R.1 and n. 2017-14.H.0
This work has made use of the INTEGRAL archive developed at the institute INAF-IASF Milano (http://www.iasf-milano.inaf.it/{\textasciitilde}ada/GOLIA.html) and it has made use of data  provided by the  High Energy Astrophysics Science Archive Research Center (HEASARC), which is a  service of the Astrophysics Science Division at NASA/GSFC and the  High Energy Astrophysics Division of the Smithsonian Astrophysical Observatory.

\bibliographystyle{mn2e} 
\bibliographystyle{mnras}
\bibliography{mybiblio.bib}

{}

\end{document}